\documentclass[aps,prb,showpacs,citeautoscript,superscriptaddress,twocolumn]{revtex4-1}

\usepackage{graphicx}
\usepackage{amsmath}
\usepackage{amssymb}
\usepackage{color}
\usepackage[toc,page]{appendix}
\usepackage{environ}
\usepackage{tikz}
\usetikzlibrary{shapes,arrows}

\newcommand{\bea}{\begin{eqnarray*}}
\newcommand{\eea}{\end{eqnarray*}}
\newcommand{\bne}{\begin{equation*}}
\newcommand{\ede}{\end{equation*}}

\newcommand{\bnen}{\begin{equation}}
\newcommand{\eden}{\end{equation}}
\newcommand{\bean}{\begin{eqnarray}}
\newcommand{\eean}{\end{eqnarray}}
\newcommand{\bnsn}{\begin{subequations}}
\newcommand{\edsn}{\end{subequations}}

\newcommand{\bna}{\begin{array}}
\newcommand{\eda}{\end{array}}
\newcommand{\bnm}{\begin{enumerate}}
\newcommand{\edm}{\end{enumerate}}
\newcommand{\bni}{\begin{itemize}}
\newcommand{\edi}{\end{itemize}}

\newcommand{\tvec}[3]{\left(\bna{c} #1 \\[1.5ex] #2 \\[1.5ex] #3 \eda\right)}

\renewcommand{\vec}[1]{\text{\boldmath{$ #1 $}}}

\newcommand{\avg}[1]{\langle #1 \rangle}

\newcommand{\ket}[1]{| #1 \rangle}
\newcommand{\bra}[1]{\langle #1 |}

\begin{document}

\title{Orbital hyperfine interaction and qubit 
dephasing in carbon nanotube quantum dots}

\author{G\'abor Csisz\'ar}
\affiliation{Institute of Physics, E\"otv\"os University, Budapest, Hungary}

\author{Andr\'as P\'alyi}
\affiliation{Institute of Physics, E\"otv\"os University, Budapest, Hungary}
\affiliation{MTA-BME Condensed Matter Research Group,
Budapest University of Technology and Economics, Budapest, Hungary}
\affiliation{Kavli Institute for Theoretical Physics China, Beijing, P. R. China}

\date{\today}

\newcommand{\Dso}{\Delta_{\rm SO}}
\begin{abstract}
Hyperfine interaction (HF) is of key importance for 
the functionality of solid-state quantum information processing, as
it affects qubit coherence and enables nuclear-spin 
quantum memories.
In this work, we complete the theory of the 
basic hyperfine interaction mechanisms (Fermi contact, dipolar,
orbital) in carbon nanotube quantum dots by providing a theoretical 
description of the orbital HF. 
We find that orbital HF induces an interaction between 
the nuclear spins of the nanotube lattice and the 
valley degree of freedom of the electrons confined in the quantum dot.
We show that the resulting nuclear-spin--electron-valley interaction 
(i) is approximately of Ising type, 
(ii) is essentially local, in the sense that 
a radius- and dot-length-independent 
atomic interaction strength can be defined, and
(iii) has an atomic interaction strength that is comparable to
the combined strength of Fermi contact and dipolar interactions.
We argue that orbital HF provides a new decoherence mechanism 
 for
single-electron valley qubits and spin-valley qubits
in a range of multi-valley materials.
We explicitly evaluate the corresponding inhomogeneous dephasing 
time $T_2^*$ for a nanotube-based valley qubit.
\end{abstract}

\pacs{73.63.Kv, 73.63.Fg, 71.70.Ej, 76.20.+q}
\maketitle



\section{Introduction}

Carbon nanotubes (CNT) provide a promising platform\cite{Kuemmeth-review,Laird-cm-review}
for 
quantum information processing\cite{Loss-divincenzo,Hanson-rmp,Hanson-naturereview,Awschalom-sciencereview}:
In non-metallic CNTs, one or a few electrons can be captured in 
an electrically defined quantum dot (QD), 
potentially allowing for coherent control of
the electrons' internal (spin and valley) degrees of freedom.

Hyperfine interaction (HF) between the nuclear spins of the lattice and
the electrons in the QD can be either a nuisance or an asset
in this context. 
On the one hand, a randomized nuclear-spin
ensemble induces decoherence
of a spin-based electronic qubit\cite{Koppens-esr,Merkulov-hyperfine,Fischer-prb-cnt}. 
On the other hand, HF is the mechanism that allows
for information transfer between the electronic state and the nuclear spins,
a critical step for utilizing nuclear spins as long-lived quantum memories\cite{Kane,Taylor-memory,Dutt,Fuchs,Pla}.
Remarkably, the abundance of nuclear spins in the CNT lattice can 
be increased (decreased) by isotopic enrichment\cite{Simon-isotopeengineering,ChurchillPRL,ChurchillNPhys} (purification) of
the spin-half $^{13}$C nuclei, which have a natural abundance of $\sim 1$\%.
The fundamental importance of HF
in these nanostructures
is also highlighted by the possibility of HF-mediated
nuclear magnetism in 
one-dimensional solids\cite{Braunecker1,Braunecker2,Scheller}
including $^{13}$C-enriched CNTs.

\begin{figure}
\includegraphics[width=0.4\textwidth]{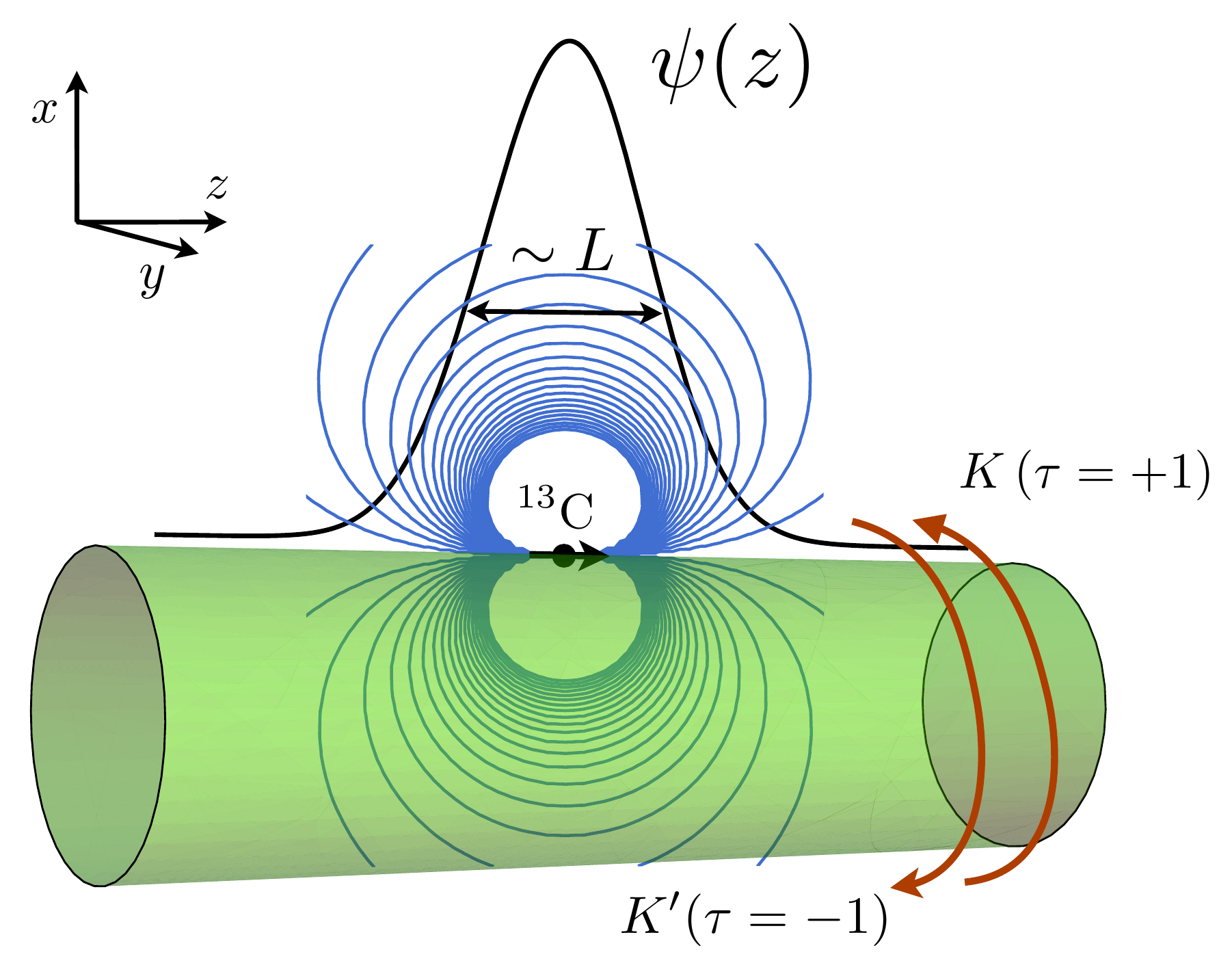}
\caption{\label{fig:CNT-dipole-coordsys} 
Carbon nanotube quantum dot with a single spin-carrying $^{13}$C nucleus. 
Arrows labeled with 
$K\left(\tau=+1\right)$ and $K'\left(\tau=-1\right)$ represent the two 
electronic valley states, moving to opposite directions around the nanotube 
circumference.
The black solid line and $\psi(z)$ represents the longitudinal
envelope-function characterizing the ground-state
of an electron confined in the QD. 
Blue lines represent the dipole magnetic field created by the 
spin of the $^{13}$C nucleus.
}
\end{figure}

Partly motivated by these attractive features,
a series of experiments were carried out with clean
CNTs, aiming to control and measure  
the spin and valley degrees of freedom of electrons confined in QDs
\cite{ChurchillPRL,ChurchillNPhys,FeiPei,Laird}.
Surprisingly, two of these experiments using $^{13}$C-enriched
samples revealed effects compatible with
an atomic HF strength that is two orders of magnitude
larger than theoretically calculated \cite{Yazyev,Fischer-prb-cnt}
and measured via nuclear magnetic resonance\cite{Pennington,Kiss}.
The resolution of this discrepancy is an open problem\cite{Laird-cm-review},
bearing strong relevance for HF-related phenomena in
CNTs.

The interesting prospects in quantum information processing and
nuclear magnetism, as well as the theory-experiment 
mismatch of the coupling strength, stimulated  
efforts \cite{Palyi-hyperfine,Reynoso1,Reynoso2,Kiss,Fuchs1,Fuchs2}
toward a more complete understanding of HF 
in carbon-based nanostructures.  
These works explore the consequences of
two out of the three basic mechanisms
of HF\cite{Abragam}, Fermi contact 
(a.k.a. isotropic) and dipolar, and exclude
the third one, orbital HF (OHF)\cite{Yafet}. 
It should be noted that the consequences of OHF
in nuclear magnetic resonance of CNTs\cite{Latil,GozeBac} 
and graphene\cite{Dora-graphenenmr} have been analyzed.

In this work, we complete the theoretical description of hyperfine effects
in CNT QDs by elucidating the role of the 
OHF.
We show that this mechanism provides an effective interaction between
a nuclear spin and the valley degree of freedom of the electron:
the simple argument (see Fig. \ref{fig:CNT-dipole-coordsys}) 
is that the binary valley quantum number $K$,
$K'$ labels electronic states circulating along the CNT
circumference in the clockwise and counter-clockwise 
direction\cite{Ajiki,Minot}, 
respectively, 
and therefore the electron has  a valley-dependent
orbital magnetic moment that
feels the dipole magnetic field created by the nuclear spin.
Using the envelope-function model (Dirac equation) for the electrons,
and focusing on the case when the longitudinal electronic wave length
$\lambda$
exceeds the nanotube radius $R$, 
we show that the resulting nuclear-spin--electron-valley interaction 
(i) is approximately of Ising type, 
(ii) is essentially local, in the sense that 
a radius- and dot-length-independent 
atomic interaction strength can be defined, and
(iii) has an atomic interaction strength that is comparable to
the combined strength of Fermi contact and dipolar interactions.
We argue that the inhomogeneous dephasing time $T_2^*$ of 
single-electron valley qubits and spin-valley qubits is affected by the
OHF, and explicitly evaluate $T_2^*$ for a valley qubit.

\section{Orbital hyperfine interaction with the electronic
valley degree of freedom}

Here, we provide an analytical
description of the OHF-mediated coupling between 
the nuclear spin of a single $^{13}$C atom residing in
a CNT QD and the valley
degree of freedom of a single electron confined to the same QD. 
To this end, the electron will be described by the canonical
envelope-function model\cite{Laird-cm-review} of CNTs (Dirac equation). 
In the terminology introduced by Yafet\cite{Yafet},
our approach describes the `long-range' part of OHF; 
the `short-range' part of OHF is shown to be absent in CNTs
in the tight-binding framework of Ref. \onlinecite{Fischer-prb-cnt}.
Our description remains qualitatively 
valid for any type of spin-carrying nucleus.

The setup and the reference frame is 
shown in Fig. \ref{fig:CNT-dipole-coordsys}.
The spin-carrying $^{13}$C nucleus is located at 
$\vec r_0  = (R,0,z_0)$.
The nuclear spin has a dipole moment $g_N \mu_N$, and therefore
it creates a vector potential 
\bnen
\label{eq:vectorpot}
\vec A (\vec r) = \frac {\mu_0 } {4\pi }  g_N \mu_N
\frac{\vec I \times (\vec r-\vec r_0)}{|\vec r - \vec r_0|^3}
\equiv \vec I \times \vec v(\vec r-\vec r_0).
\eden
Here,
$g_{N} \approx 1.41$ is the g-factor of the $^{13}$C
nucleus\cite{Fischer-prb-cnt}, 
and $\mu_{N}\approx5.05\times 10^{-27}$ J/T
is the nuclear magneton. 
The nuclear spin vector operator $\vec I$
is represented by one half
times the vector of Pauli matrices.  

Technically, OHF between the nuclear spin and the electron arises
because the vector potential $\vec A$ created by 
the nuclear spin enters the kinetic term of the envelope-function
Hamiltonian via $\vec p \mapsto \vec p + e \vec A$. 
The kinetic  term, describing an electron in the conduction
or valence band, reads\cite{Laird-cm-review}
\bean
\label{eq:efh}
H_0 +H_{\rm OHF} &=& v_{\rm F} \left\{
		\tau_3 \sigma_1 \left[p_c+e\,  A_c(\vec r)\right]  + \sigma_2 \left[p_t+e\,  A_t(\vec r)\right] 
	\right\},\nonumber \\ 
\eean
where $c$ ($t$) is the circumferential (longitudinal) 
coordinate on the surface of the CNT, 
$\sigma_{1,2}$ are sublattice Pauli matrices, 
and $H_{\rm OHF}$ is defined as the parts of the rhs that
contain the vector potential. 
Note that our choice of the reference frame 
(Fig. \ref{fig:CNT-dipole-coordsys})
allows us to use $t$ and $z$ interchangeably. 
In Eq. \eqref{eq:efh}, we introduced the circumferential and longitudinal
 projections
of the vector potential, 
$A_c(\vec r) = \vec{\hat c}(c) \cdot \vec A(\vec r)$ and
$A_t(\vec r) = \vec{\hat t} \cdot \vec A(\vec r)$, 
respectively,
where 
$\vec{\hat c}(c) = (-\sin \frac c R, \cos \frac c R,0)$,
$\vec{\hat t} = (0,0,1)$, and
$\vec r \equiv \vec r(c,t) = (R \cos \frac c R, R\sin \frac c R,t)$.
While $p_c$ is the circumferential momentum quantum number set 
by the periodic boundary condition along the CNT circumference, 
$p_t$ is the longitudinal momentum operator. 
The form \eqref{eq:efh} of the Hamiltonian is valid for any chirality;
here we focus on CNTs with a finite gap (i.e., $p_c \neq 0$) allowing
for electrostatic QD confinement.

Using Eqs. \eqref{eq:vectorpot} and \eqref{eq:efh}, 
the OHF Hamiltonian
can be written as
\bean
\label{eq:ohfvalley}
H_{\rm OHF} &=& e v_{\rm F} \tau_3 \sigma_1 
\varepsilon_{\alpha \beta \gamma} I_\alpha v_\beta(\vec r- \vec r_0) \hat{c}_\gamma,
\eean
where $\varepsilon_{\alpha \beta \gamma}$ is the Levi-Civita symbol,
the Einsten summation convention is used,
and the valley-independent term have been omitted as it is 
irrelevant for valley dynamics. 

For simplicity, we assume $p_c >0$, and anticipate that a 
sign change in $p_c$ implies a sign change of the coupling constants
$C_\alpha$ (defined below). 
Then, an electronic low-energy energy eigenstate in 
valley $\tau \in (K,K') \equiv (+1,-1)$ of the conduction band
of the electrostatically defined QD is approximately described by the 
four-component spinor envelope function
\bean
\label{eq:env}
\Psi_{\tau}(c,t)  &=& 
\ket{\tau} \otimes \ket{\chi } \otimes 
\frac{e^{i \tau (p_c/\hbar) c}}{\sqrt{2\pi R}} \psi(t),
\eean
where 
$\ket{\tau=+1} = (1,0)^T$ or
$\ket{\tau=-1} = (0,1)^T$ represent the valley state,
$\ket{\chi} = (1,1)^T/\sqrt{2}$ characterizes the sublattice amplitudes
at the bottom of the conduction band, 
and 
$\psi(t)$ is the longitudinal envelope function of the electron.
The normalization condition
$ \int_{-\infty}^{\infty} dt \int_{0}^{2\pi R} dc
\Psi^\dag(c,t) \Psi(c,t) = 1$ demands
$\int_{-\infty}^{\infty} dt |\psi(t)|^2 = 1$. 
Note that by writing the envelope function $\Psi_\tau(c,t)$ as
a product of a circumferential and longitudinal component in 
Eq. \eqref{eq:env}, we implicitly assumed that the 
confinement potential is longitudinal (i.e., independent of $c$). 

The effective Hamiltonian describing the nuclear-spin--electron-valley
interaction  is  obtained via first-order 
degenerate perturbation theory, i.e., by projecting $H_{\rm OHF}$
to the two-dimensional subspace spanned by
$\Psi_K$ and $\Psi_{K'}$:
\bean
\label{eq:eff}
H_{\rm OHF}^{\rm (eff)} \equiv P H_{\rm OHF} P =
\frac 1 2 \tau_3 \sum_{\alpha=x,y,z} C_\alpha I_\alpha,
\eean
where Eqs. \eqref{eq:ohfvalley}, \eqref{eq:env},
and $P \equiv \ket{\Psi_K}\bra{\Psi_{K}} + \ket{\Psi_{K'}}\bra{\Psi_{K'}}$ 
were used, 
$\tau_3$ has been redefined as
$\tau_3 \equiv \ket{\Psi_K}\bra{\Psi_K} - \ket{\Psi_{K'}}\bra{\Psi_{K'}}$,
and
\bean
\label{eq:ci}
C_\alpha(z_0) =
\frac{e v_{\rm F}}{\pi R}
\int_{-\infty}^{\infty} dt |\psi(t)|^2
\int_{0}^{2\pi R} dc \, 
\epsilon_{\alpha \beta \gamma} v_\beta(\vec r - \vec r_0) \hat c_\gamma.
\nonumber \\
\eean
Note that the last  integral ($\int dc \dots$) is proportional to the
magnetic flux that is piercing the circular cross section of the CNT 
at height $t$ in the presence of a (classical)
nuclear spin that is aligned with axis $\alpha$.

\begin{figure}
\includegraphics[width=0.45\textwidth]{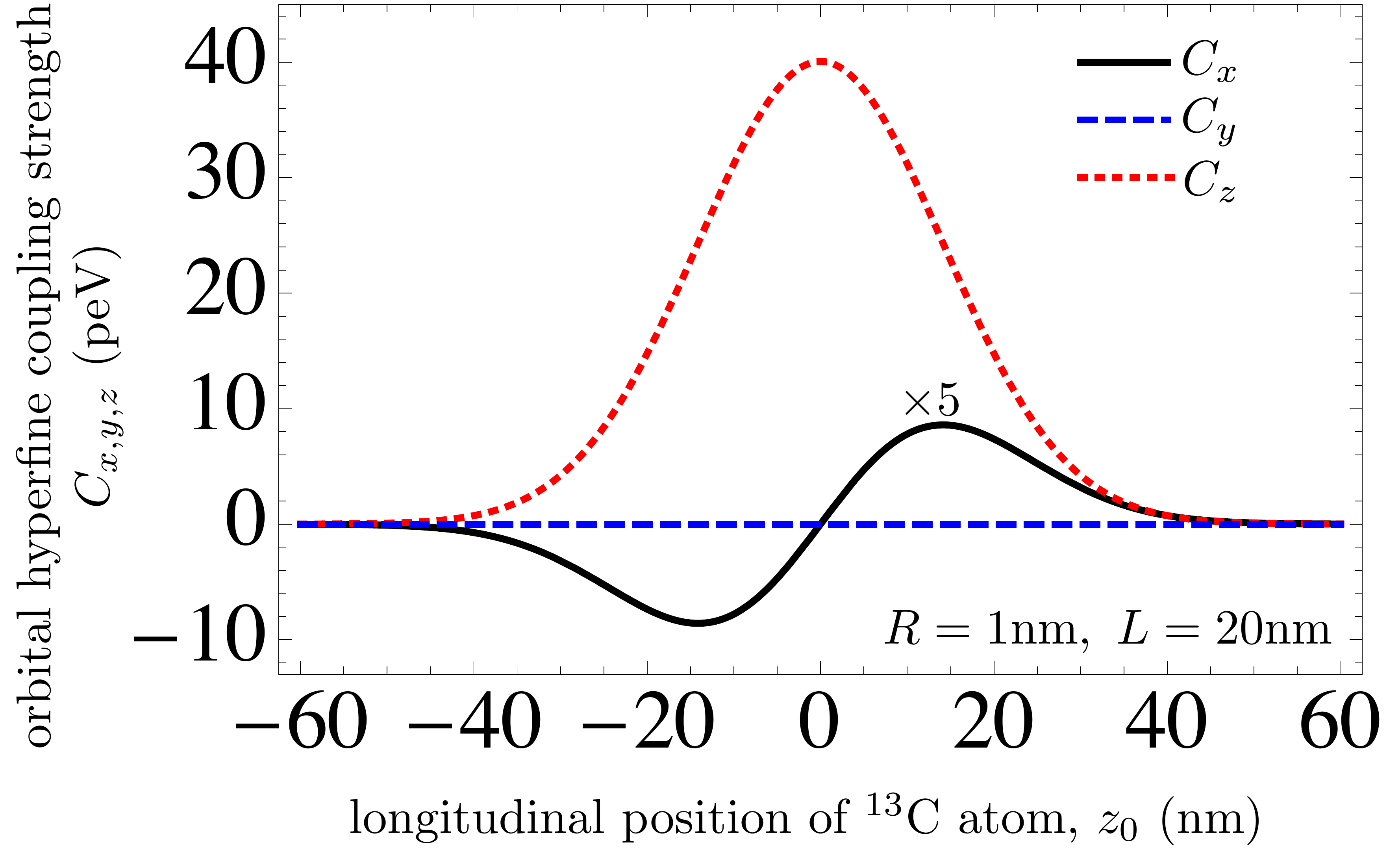}
\caption{\label{fig:couplingstrengths} 
Orbital hyperfine coupling strengths as functions 
of the nuclear-spin position in a carbon nanotube quantum dot.
The plot is based on Eq. \eqref{eq:result} and corresponds
to the case of a Gaussian longitudinal envelope function
[Eq. \eqref{eq:gaussian}].
Note that $C_x$ was multiplied by 5.
Axes $x$, $y$, $z$ are defined in Fig. \ref{fig:CNT-dipole-coordsys}.}
\end{figure}

Now we evaluate the coupling strengths in
Eq. \eqref{eq:ci}, in the case where the length
scale $\lambda$ of the spatial variation of the longitudinal envelope function
$\psi(z)$ exceeds the radius $R$ of the CNT. 
This is the relevant case for those experiments that are 
done with few electrons confined in a $\sim 100$ nm long QD
in a CNT with radius $R\sim 1$nm.
For this case, we will show that 
\begin{subequations}
\label{eq:result}
\bean
\label{eq:resultx}
C_x(z_0) &\approx& -
2 e v_{\rm F} 
\frac{\mu_0 g_N \mu_N}{4\pi}
[\psi^*(z_0) \psi'(z_0) +c.c.]
\\
\label{eq:resulty}
C_y (z_0) &=& 0,\\
\label{eq:resultz}
C_z(z_0) &\approx  &
2 e v_{\rm F} \frac{\mu_0 g_N \mu_N}{4\pi}
\frac 1 R |\psi( z_0)|^2,
\eean
\end{subequations}
where $\psi'$ is the derivative of $\psi$ with respect to 
the longitudinal coordinate. 
Equation \eqref{eq:eff} together with Eq. \eqref{eq:result} form
the central result of this work. 
The dependence of the three coupling strengths $C_{x,y,z}$ on
the longitudinal position $z_0$ of the nuclear spin is
shown in Fig. \ref{fig:couplingstrengths},
for the case of a CNT with radius $R=1$ nm and 
a Gaussian longitudinal envelope function
\bean
\label{eq:gaussian}
\psi(t) = \frac{1}{\pi^{1/4} \sqrt{L}} e^{-\frac{t^2}{2 L^2}}
\eean
with $L=20$ nm.

An interpretation of Eq. \eqref{eq:result} is as follows. 
If the nuclear spin is aligned with the CNT axis, then it induces
an energy splitting $C_z$ between the two valley states 
(i.e., a \emph{valley splitting})
of the electron. 
If the nuclear spin is aligned radially, then it induces
a valley splitting $C_x$, which is 
typically much smaller than in the former case,
because $(\partial_z \psi)(z_0)\sim\psi(z_0)/\lambda$ is much smaller than 
$\psi(z_0)/R$ due to the assumed length-scale mismatch $R \ll \lambda $.
Finally,  if the nuclear spin is aligned orthoradially (i.e., perpendicular to
the axial and radial directions) then it 
does not induce valley splitting. 
Alternatively, Eqs. \eqref{eq:eff} and \eqref{eq:result} can be interpreted
in terms of an effective magnetic field
that acts on the nuclear spin and determined by the the valley state
of the electron. 
This effective magnetic field 
has an axial as well as a much smaller radial
component, and it has no orthoradial component.

\begin{figure}
\includegraphics[width=0.4\textwidth]{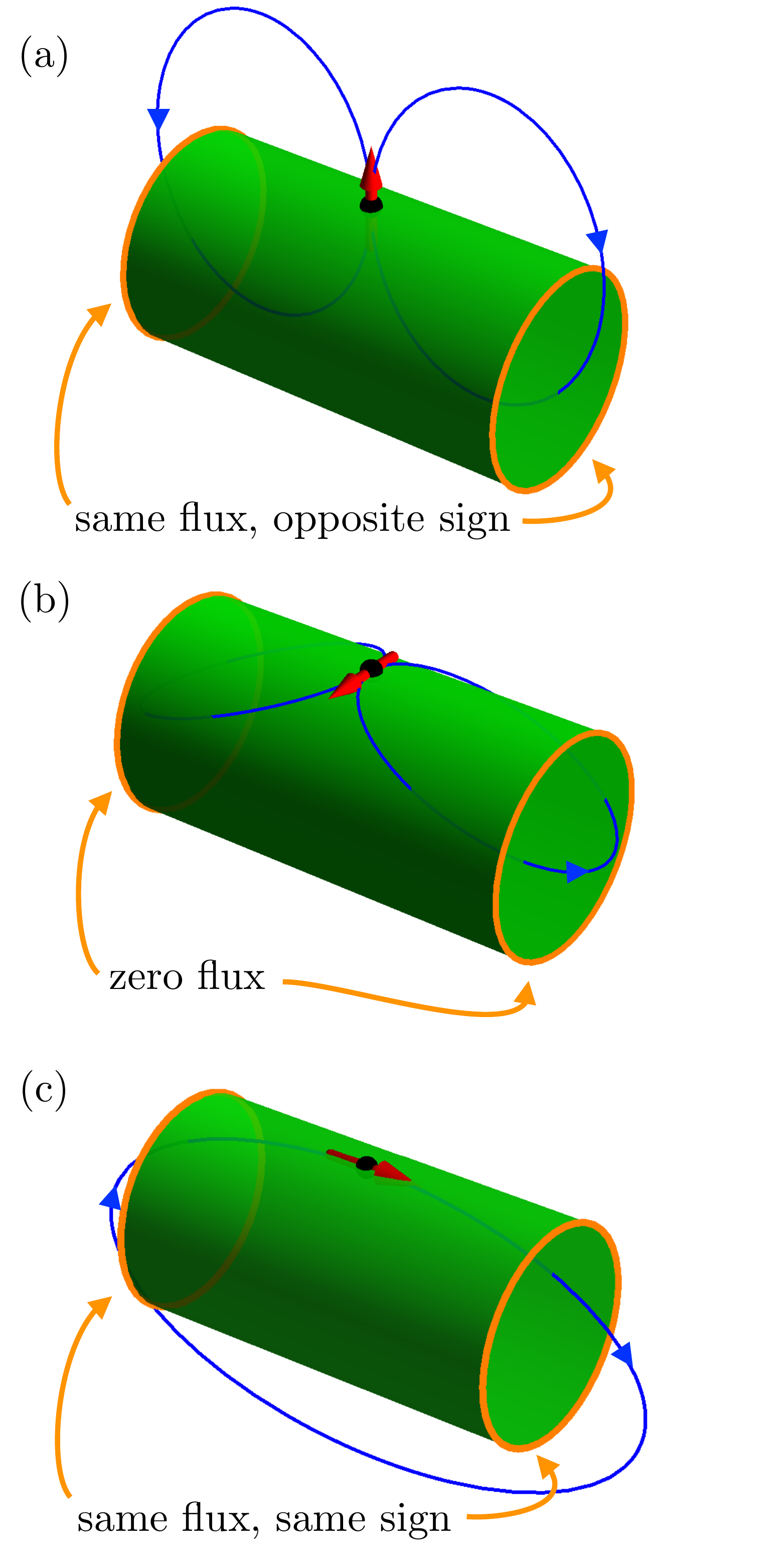}
\caption{\label{fig:flux} 
Nuclear-spin-induced magnetic field lines and 
magnetic fluxes. 
Black dot (red arrow) represents the position (alignment) of 
the nuclear spin.
Blue lines are magnetic field lines, and the blue arrowheads
indicate their directionality. 
Orange circles represent two symmetrically positioned
circular cross sections of the tube. 
(a) For a radially aligned nuclear spin, the fluxes piercing the 
two  cross sections   (orange) are
identical in magnitude but differ in sign. 
(b) For an orthoradially aligned nuclear spin, 
the flux piercing both cross sections (orange) are zero.
(c) For an axially aligned nuclear spin, the fluxes piercing the 
cross sections (orange) are identical in magnitude and sign. 
}
\end{figure}

The three coupling strengths $C_{x,y,z}$ expressed in 
Eq. \eqref{eq:result} have qualitatively different dependencies
on the longitudinal envelope function $\psi$. 
A simple understanding of these differences is gained 
using
the relation between the coupling strengths $C_\alpha$
and the magnetic fluxes piercing the CNT cross sections,
discussed after Eq. \eqref{eq:ci}.
Figure \ref{fig:flux} displays characteristic magnetic field lines piercing
circular cross sections of the CNT that are positioned symmetrically
with respect to the nuclear spin position, for three different alignments
of the nuclear spin.
For a radially aligned nuclear spin (Fig. \ref{fig:flux}a), the fluxes piercing the 
two cross sections of the tube (orange) are
identical in magnitude but differ in sign. 
Therefore, a homogeneous longitudinal envelope function $\psi(t)$
would imply a zero coupling strength $C_x$, since the flux contributions
of the two cross sections would cancel each other in the
$\int dt$ integral of Eq. \eqref{eq:ci}.
The inhomogeneity of $\psi(t)$, i.e., the finiteness of $\psi'$, 
prevents this cancellation, and allows for a finite coupling strength $C_x$;
this is reflected by the dependence 
$C_x \propto \psi'$ of Eq. \eqref{eq:resultx}.
For an axially aligned nuclear spin (Fig. \ref{fig:flux}c), the fluxes piercing the 
two circular cross sections (orange) of the CNT
are identical in sign (and also in magnitude), hence
the cancellation affecting $C_x$ is not relevant for $C_z$. 
Finally, for an orthoradially aligned nuclear spin (Fig. \ref{fig:flux}b), 
the magnetic flux piercing each of the two circular cross sections (orange)
of the tube is zero, explaining Eq. \eqref{eq:resulty}.

The results \eqref{eq:eff} and \eqref{eq:result} have the following implications. 

(i) In the considered range $\lambda \gg R$, 
the OHF-induced nuclear-spin--electron-valley interaction
is essentially of Ising type, $\propto \tau_3 I_z$.
The correction of the form $\propto \tau_3 I_x$ is small since $C_x \ll C_z$.
Note that the coupling strength $C_x$ might gain importance in
the case $\lambda \sim R$, e.g., in
ultrashort CNT QDs\cite{Sun-ultrasmall,Island-ultrasmall,Petit-ultrasmall}, 
or in QDs where the electron occupies a highly excited,
short-wavelength longitudinal mode. 

(ii) Even though OHF is long-range in principle, 
our leading-order result \eqref{eq:resultz} suggests that it is
essentially local under our assumptions, 
since the strength of the resulting nuclear-spin--electron-valley
interaction is determined by the value
of the electronic envelope function at the position
of the nucleus.
In other words, the result \eqref{eq:resultz} affirms that
for practical purposes, the envelope-function Hamiltonian
$H_{\rm OHF}$ can be replaced by
\bnen
\label{eq:local}
\tilde H_{\rm OHF} = e v_{\rm F} \mu_0 g_N \mu_N 
 \delta(c-c_0) \delta(t-t_0) \frac {\tau_3}{2} I_z,
\eden
since $P H_{\rm OHF} P = P \tilde H_{\rm OHF} P$. 
Therefore, an \emph{atomic coupling strength} $A$ of the
OHF can be defined via
\bean
\label{eq:Adef}
\tilde H_{\rm OHF} = 
\frac{\Omega_{\rm cell}}{2} A 
\delta(c-c_0)\delta(t-t_0)
\frac{\tau_3}{2} I_z
\eean
in analogy with, e.g., the atomic coupling strength of Fermi contact HF
in GaAs, see Eq. (2) of Ref. \onlinecite{Merkulov-hyperfine}.
Here $\Omega_{\rm cell} \approx 5.24$ \AA$^2$ is the area of the
graphene unit cell.
The atomic coupling strength  $A$ can be deduced from 
 Eqs. \eqref{eq:local} and \eqref{eq:Adef}:
\bean
\label{eq:estimate}
A\approx \frac{2 ev_{\rm F} \mu_0 g_N \mu_N}{\Omega_{\rm cell}}
\approx
0.34 \mu{\rm eV},
\eean
where $v_{\rm F}=10^6$ m/s was assumed\cite{Laird-cm-review}.

(iii) The estimated atomic coupling strength \eqref{eq:estimate}
of the OHF-induced nuclear-spin--electron-valley
interaction 
is comparable to the atomic coupling strength of the combined
Fermi contact and dipolar
 spin HF\cite{Fischer-prb-cnt,Yazyev}.
This result has the following consequences.
(1) In order to provide an accurate assessment of any property or 
functionality of a CNT-based electronic spin-valley qubit\cite
{FlensbergMarcus,FeiPei,Laird,Szechenyi-maximalrabi,Osika},
the Fermi contact, dipolar and orbital contributions should be treated
on an equal footing.
(2) Nuclear spins in a CNT QD will induce inhomogeneous dephasing
of an electronic valley qubit on a time scale similar to the
inhomogeneous dephasing time of a spin qubit or spin-valley qubit. 
We present a detailed analysis of the latter point
in Sec. \ref{sec:valleyqubit}.

For completeness we also provide the OHF Hamiltonian 
describing 
the nuclear-electron spin-valley interaction in the
presence of more than one nuclear spins $\vec I_k$.
Here, $k \in 1\dots N$ where $N$ is the number of atoms interacting
with the electron in the QD, and 
$\vec I_k$ is the spin-half nuclear spin operator if site $k$ has
a $^{13}$C atom and zero otherwise. 
The $k$th atom is assumed to be located
at the position specified by the circumferential $c_k$ and longitudinal $t_k$
coordinates, 
corresponding to the real-space position 
$\vec r_k \equiv (x_k,y_k,z_k) =  (R \cos(c_k/R), R\sin(c_k/R),t_k)$.
The effective OHF Hamiltonian then reads
\bean
\label{eq:manynuclearspins}
H_{\rm OHF}^{\rm (eff)} = 
\frac 1 2 \vec \tau \sum_{k=1}^{N}
M_k \vec I_k,
\eean
where $M_k$ is the $3\times 3$ local orbital hyperfine tensor 
\bean
M_k = 
\left( \begin{array}{ccc}
	0 & 0 & 0 \\
	0 & 0 & 0 \\
	C_x(t_k) \cos(c_k/R) & C_x(t_k) \sin(c_k/R) & C_z(t_k)
\end{array}\right),
\eean
and $C_x$ and $C_z$ are given in 
Eq. \eqref{eq:result}.

This Section is concluded by proving Eq. \eqref{eq:result}.
First, we prove Eq. \eqref{eq:resultx} by evaluating
$C_x$ using Eq. \eqref{eq:ci},
the definition of $\vec v$ via Eq. \eqref{eq:vectorpot}, and
the definition of $\vec{\hat c}$ given below Eq. \eqref{eq:efh}.
After introducing the dimensionless
quantities $\varphi = c/R$ and $\zeta = \frac{t-z_0}{R}$, 
we find
\bean
\label{eq:cxintegral}
C_x(z_0) &=& -
\frac{e v_{\rm F} }{\pi R} \frac{\mu_0 g_N \mu_N}{4\pi}
\int_{-\infty}^{\infty} d \zeta |\psi(z_0 + R \zeta)|^2
 \nonumber \\
&\times &
\int_{0}^{2\pi} d\varphi 
 \frac{\zeta \cos \left(\varphi\right)}{\left(2 -2  \cos \left(\varphi\right) +\zeta^2\right)^{3/2}}
\eean
Note that although the integrand is singular at the position of the nuclear spin,
i.e., for $(c,t)=(0,z_0)$, that is, for $(\varphi,\zeta) = (0,0)$, 
the integral does converge, similarly to the case of 
OHF in graphene\cite{Dora-graphenenmr}.
Importantly, the integrand in the second line of Eq. \eqref{eq:cxintegral}
decays for $\zeta \gg 1$ as $\sim 1/\zeta^2$, suggesting that the integral
is dominated by the range $|\zeta| \lesssim 1$, that is, 
by the spatial range $|z-z_0| \lesssim R$.
Since this spatial range is narrow in comparison with the length scale 
$\lambda$
characterizing the spatial variation of the 
longitudinal envelope function $\psi$, 
we can estimate the value of the integral by 
expanding the term $|\psi |^2$ term up to first order in $\zeta$:
$|\psi(z_0+R\zeta)|^2 \approx |\psi(z_0)|^2+ R \zeta [\psi^*(z_0) \psi'(z_0)+c.c.]$.
The integral containing the zeroth-order term vanishes as its
integrand is an odd function of $\zeta$. 
The integral containing the first-order term is finite though, 
providing the estimate Eq. \eqref{eq:resultx} above. 
Technically, the $C_y = 0$ result in Eq. \eqref{eq:resulty} follows from the 
fact that the $\alpha=y$ integrand in Eq. \eqref{eq:ci}
is an antisymmetric function of $c \in [0,2\pi R]$ 
with respect to $c=\pi R$. 
Finally, the derivation of Eq. \eqref{eq:resultz} is similar to that of 
Eq. \eqref{eq:resultx}.
The difference here is that the integral will be dominated
by the zeroth-order term in the $\zeta$ expansion
of $|\psi(z_0 + R\zeta)|^2$, 
hence higher-order terms
can be neglected.

\section{Inhomogeneous dephasing of a valley qubit}
\label{sec:valleyqubit}

HF leads to information loss via decoherence for 
spin qubits
in conventional semiconductors\cite{Merkulov-hyperfine,Coish-review} 
as well as in carbon nanostructures\cite{Fischer-prb-cnt,Fuchs1,Fuchs2}.
One form of decoherence is inhomogeneous dephasing, 
which arises due to a nuclear-spin-induced random component in the
qubit's Larmor frequency, 
and is characterized by the time scale $T_2^*$, usually called
the inhomogeneous dephasing time. 
To our knowledge, the $T_2^*$ of valley 
qubits\cite{Recher-graphenering,Palyi-valley-resonance,Wu-valleyqubit,CulcerPRL,Rohling-spinvalley}
due to HF has not yet been investigated. 
Based on the result of the previous Section, here we evaluate 
$T_2^*$ for a valley qubit formed in the conduction-band 
ground state of a CNT QD, as a function 
of the CNT radius $R$, the QD length $L$, and the 
abundance $\nu \in [0,1]$ of $^{13}$C atoms.

We assume that the valley states of the single electron
forming the valley qubit are energy-split ($\hbar \omega_L$) by
a longitudinal magnetic field and/or spin-orbit interaction,
and the qubit is tuned far away from the $K$-$K'$ 
anticrossing\cite{Bulaev,Kuemmeth}
caused by valley mixing. 
The valley qubit is prepared in a superposition state
$\Psi(t=0) = \frac{1}{\sqrt{2}} \left(\ket{\Psi_K} + \ket{\Psi_{K'}}\right)$.
It interacts with the nuclear spin bath that is completely 
disordered to a good approximation. 
The  Hamiltonian reads
\bean
H = 
\frac{1}{2} \hbar \omega_L \tau_3 + H_{\rm OHF}^{\rm (eff)},
\eean
where
$H_{\rm OHF}^{\rm (eff)}$ is given in Eq. \eqref{eq:manynuclearspins},
and we neglect $C_x$.

Following Merkulov et al.\cite{Merkulov-hyperfine},
we disregard the slow dynamics of the nuclear spin bath,
and describe the nuclear spins as being frozen 
during the time evolution of the electronic valley state.
The influence of the nuclear spin ensemble is expressed via
the HF-induced random correction 
$\delta \omega = \frac 1 \hbar \sum_{k,\alpha} M_{k;3,\alpha} I_{k,\alpha}
\approx \frac 1 \hbar \sum_{k} C_{z}(z_k) I_{k,z}$
of the valley Larmor frequency $\omega_L$.
In the presence of many nuclear spins, 
the correction $\delta \omega $ can be regarded as a 
Gaussian random variable with the following
mean and variance:
\bean
\avg{\delta \omega} &\equiv& \frac 1 \hbar 
	\langle \sum_k C_z(z_k) I_{k,z} \rangle= 0,\\
\sigma^2 &\equiv & \avg{(\delta \omega)^2} =
\frac{\nu}{4} \frac 1 \hbar \sum_k C^2_z(z_k).
\eean
Here the average $\avg{.}$ refers to both ensemble 
averaging for the nuclear spin states as well as disorder averaging for
the possible spatial configurations of the spin-carrying nuclei. 
Correspondingly, we used
$\avg{I_{k,\alpha}} = 0$ and $\avg{I_{k,\alpha} I_{k',\alpha'}} = \frac \nu 4 \delta_{kk'} \delta_{\alpha\alpha'}$,
and $\nu$ is the abundance of spin-carrying nuclei.

The  polarization vector of the valley qubit 
in the initial state $\Psi(0)$ 
is $\vec p \equiv \bra{\Psi(0)} \vec \tau \ket{\Psi(0)} = (1,0,0)^T$.
A straightforward calculation\cite{Merkulov-hyperfine}
shows that the time evolution of the 
valley polarization $\vec p(t)$,
averaged for the random nuclear-spin configurations, reads
\bean
\avg{\vec p}(t) \equiv
\bra{\Psi(t)}
	\vec \tau
\ket{\Psi(t)}
=
\tvec{\cos(\omega_L t)}{\sin (\omega_L t)}{0}
e^{-(t/T_2^*)^2},
\eean
where 
\bean
\label{eq:t2star}
T_2^* =
 \sqrt{2}/\sigma = 
 \frac{1}{\sqrt \nu}
 \frac{2 \sqrt{2} \hbar }{\sqrt{ \sum_k C_z^2(z_k)}}.
\eean

For a box-type longitudinal envelope
function, i.e., if $\psi(z) = 1/\sqrt{L}$ within
 a QD of length $L$,  we find
\bean
\label{eq:t2starBox}
T_2^* \approx \frac{1}{\sqrt \nu} 2^{3/2}\sqrt{\pi}
\frac{\sqrt{\Omega_{\rm cell}} \hbar }{ev_{\rm F} \mu_0 g_N \mu_N}\sqrt{LR},
\eean
where Eqs. \eqref{eq:resultz} and \eqref{eq:t2star} were used.
For a natural (non-isotope-enriched)
CNT QD containing $N=6\times 10^5$ atoms, 
we estimate $T_2^* \approx 266\, \mu$s, comparable
to the theoretically estimated spin dephasing time\cite{Fischer-prb-cnt}.
This is not surprising, regarding that the  orbital HF atomic
coupling strength estimated in the previous Section was also 
comparable to the spin HF (combined Fermi contact and dipolar)
atomic coupling strength. 

For the Gaussian longitudinal envelope function defined in 
Eq. \eqref{eq:gaussian}, 
which provides a more realistic description of the ground-state
orbital of a QD with parabolic electrostatic confinement, we
find the same parameter dependence as for the box-model wave function,
with slightly different prefactors: 
\bean
\label{eq:t2starGaussian}
T_2^* \approx \frac{1}{\sqrt{\nu}} 2^{7/4}\pi ^{3/4} 
\frac{ \sqrt{\Omega_{\rm cell}} \hbar}
{ e v_{\rm F} \mu_0 g_N \mu_N 
} \sqrt{LR}.
\eean
Note that if the relatively small coupling strength $C_x$ corresponding to 
a radially aligned nuclear spin is taken into account, then the 
inhomogeneous dephasing time is $\left(1-\frac{R^2}{2L^2}\right)$ times
the rhs of Eq. \eqref{eq:t2starGaussian}, i.e., 
the correction due to $C_x$ is second order in the small quantity $R/L$. 
The coupling strength $C_x$ might gain importance and 
significantly contribute to the $T_2^*$ in 
ultrashort CNT QDs\cite{Sun-ultrasmall,Island-ultrasmall,Petit-ultrasmall},
where $L\sim R$, 
or for electrons that occupy a highly excited longitudinal mode of 
a QD.

The interpretation of the results \eqref{eq:t2starBox} and
\eqref{eq:t2starGaussian} is straightforward. 
An increasing $^{13}$C abundance $\nu$ leads to shorter
$T_2^*$,  and the inverse-square-root dependence on $\nu$
originates from
the completely randomized character of the nuclear spin bath. 
The inverse linear dependence of $T_2^*$ on the parameters setting the
interaction strength ($e$, $v_F$, $\mu_0$, $g_N$, $\mu_N$) 
and the square-root dependence on the geometrical parameters
$R$ and $L$ are natural consequences of the parametric
dependencies of the orbital HF coupling strengths in
Eq. \eqref{eq:result}.

\section{Discussion}

(1) A natural consequence of our results is that 
the nuclear-spin--electron-valley interaction arising from OHF
contributes to the dephasing of spin-valley qubits in 
carbon nanotubes\cite{FlensbergMarcus,Laird}.
To our knowledge, this contribution has not been analyzed to date. 
As the strength of this interaction is comparable to the nuclear-spin--electron-spin\cite{Yazyev,Fischer-prb-cnt} 
and nuclear-spin--electron-valley\cite{Palyi-hyperfine} interactions arising from 
the Fermi contact and dipolar mechanisms, it is
expected that the hyperfine-limited inhomogeneous dephasing time of a 
spin-valley qubit has a scale similar to that of the
spin qubit (see Ref. \onlinecite{Fischer-prb-cnt}) and the valley qubit 
(see Sec. \ref{sec:valleyqubit}). 
A detailed calculation of $T_2^*$ of the spin-valley qubit, 
which could quantify, e.g., 
the dependence of $T_2^*$ on the direction of the homogeneous 
magnetic field, is yet to be done. 

(2) We emphasize that valley-qubit coherence might be affected 
by mechanisms other than OHF.
For example, spin-independent potential disorder, similarly to the case of 
silicon-based heterostructure QDs\cite{Friesen-valleyorbit,Culcer-interfaceroughness,Gamble}, makes 
the valley qubit susceptible to electric fields\cite{Palyi-valley-resonance},
including electrical fluctuations caused by phonons or nearby electrodes.
In addition, if a CNT valley qubit is tuned by an axial magnetic field 
to the $K$-$K'$ anticrossing (e.g., to 0.11 T in Fig. 2e of Ref. \onlinecite{Kuemmeth}), then valley mixing due to Fermi contact and dipolar
HF\cite{Palyi-hyperfine} can also induce qubit dephasing.
Exploring the competition and interplay of various 
valley-qubit decoherence mechanisms is an interesting future direction. 

(3) Importantly, our present effort, which completes
the theoretical description of the basic hyperfine mechanisms for CNT QDs,
does not explain the comparatively strong 
hyperfine coupling strength deduced from the
experiments of Refs. \onlinecite{ChurchillPRL} and \onlinecite{ChurchillNPhys}.

(4) HF between electronic spin qubits and
nuclear spins can be harmful, as described above, from the
quantum information perspective. 
It can also be an asset though: 
in principle, nuclear spins can be used as long-lived
quantum memories\cite{Taylor-memory,Dutt,Fuchs,Pla}, 
and information transfer between the electronic
and nuclear degrees of freedom can be mediated by 
HF.
Furthermore, enrichment (purification) of the $^{13}$C abundance is 
a feasible way\cite{Simon-isotopeengineering,ChurchillPRL,ChurchillNPhys} to enhance (suppress) hyperfine effects. 

(5) In certain inversion-symmetry-broken 
two-dimensional multi-valley materials, 
such as monolayer transition-metal
dichalcogenides\cite{DiXiao-tmdc} 
or gapped graphene\cite{DiXiao-valley}, the electronic states acquire a
finite valley-dependent magnetic moment.
This magnetic moment is inherently coupled with
the nuclear spins of the crystal lattice via OHF;
therefore, if an electron is confined in a QD in these 
materials\cite{Kormanyos,Recher}, then its
operation as  a valley qubit or as a spin-valley qubit will be influenced by
the OHF-induced nuclear-spin--electron-valley interaction 
in a similar fashion as in a CNT. 
The OHF and its consequences 
in such two-dimensional materials are yet to be explored.

In conclusion, we have shown that orbital hyperfine interaction couples 
the nuclear spins residing in a carbon nanotube quantum dot and the 
valley degree of freedom of the electron confined in the quantum dot.
We provided a quantitative analysis of this interaction, 
and found that it is essentially a local Ising-type interaction, 
which is as strong as the nuclear-spin--electron-spin hyperfine interactions 
(Fermi contact and dipolar). 
As an application, we
evaluated the hyperfine-limited inhomogeneous dephasing time of 
a single-electron valley qubit, which was found to be in the $\sim 100 \mu$s 
range, similar to theoretical estimates for CNT-based spin qubits, 
but much longer than the measured $T_2^*$ of single-electron spin-valley 
qubits.


\begin{acknowledgments}
We thank P. Boross, B. D\'ora, P. Nagy,
M. Rudner,  F. Simon, \'A. Szabados and G. Tichy for useful discussions.
We acknowledge funding from 
the EU Marie Curie Career Integration Grant CIG-293834 (CarbonQubits),
the OTKA Grant PD 100373, and the EU ERC Starting Grant 
CooPairEnt 258789.
A.~ P.~ is supported by the 
J\'anos Bolyai Scholarship of the Hungarian Academy of Sciences.
\end{acknowledgments}




\bibliography{orbital-hyperfine}

\end{document}